\documentclass[aps,showpacs,prl,twocolumn,floatfix,superscriptaddress,floatfix]{revtex4}
\usepackage[english]{babel}
\usepackage{graphicx}
\usepackage{color}
\usepackage[utf8]{inputenc}
\usepackage{pstricks,pst-grad,color}
\usepackage{graphicx,amssymb}
\usepackage{amsmath}
\usepackage{amssymb}
\usepackage{bbold}
\usepackage{placeins}

\begin{document}

\title{Strong enhancement of the Edelstein effect in $f$-electron systems}

\begin{abstract}
The Edelstein effect occurring in systems with broken inversion symmetry generates a spin polarization when an electric field is applied, which is most advantageous in spintronics applications. Unfortunately, it became apparent that this kind of magnetoelectric effect is very small  in semiconductors. We here demonstrate that correlation effects can strongly enhance the magnetoelectric effect. Particularly, we observe a strong enhancement of the Edelstein effect in $f$-electron systems close to the coherence temperature, where the $f$-electrons change their character from localized to itinerant. We furthermore show that this enhancement can be explained  by a coupling between the conduction electrons and the still localized $f$-electrons.
\end{abstract}

\author{Robert Peters}
\email[]{peters@scphys.kyoto-u.ac.jp}
\affiliation{Department of Physics, Kyoto University, Kyoto 606-8502, Japan}

\author{Youichi Yanase}
\affiliation{Department of Physics, Kyoto University, Kyoto 606-8502, Japan}

\newcommand*{\tran}{^{\mkern-1.5mu\mathsf{T}}}
\newcommand{\1}{\mbox{1}\hspace{-0.25em}\mbox{l}}
\date{\today}



\maketitle

\section{Introduction}
 Spin-orbit interaction, which leads to a coupling between the spin of an electron and its momentum, provides the possibility to manipulate the spin polarization of a material by applying electric fields as desired for spintronics. Particular interesting are lattices without inversion symmetry, where the antisymmetric spin-orbit coupling
leads to fascinating transport properties\cite{Manchon2015} such as the anomalous Hall effect\cite{PhysRev.95.1154,PhysRev.112.739,Kato2007}, the spin Hall effect\cite{Murakami1348,PhysRevLett.92.126603}, and magnetoelectric (ME) effects\cite{Ivchenko1978,Levitov1985,Edelstein1990,PhysRevLett.99.226601,PhysRevB.78.125327,Chernyshov2009,PhysRevB.78.212405,PhysRevB.80.134403}. The latter leads to a spin polarization without an applied magnetic field when an electric current flows, which has been also confirmed in experiments\cite{PhysRevLett.93.176601,Sih2005,PhysRevLett.96.186605,PhysRevLett.97.126603}.
 Thus, the spin polarization could be controlled by electric fields, which would be a tremendous 
advantage for memory storage devices\cite{Chappert2007}. 
However, the ME effect in semiconductors with antisymmetric spin-orbit interaction is usually small, so that it  cannot be effectively used in spintronic devices.     

An analysis using Fermi liquid theory has shown that in interacting systems without inversion symmetry the ME effect can be enhanced\cite{Fujimoto2007a,Fujimoto2007b,Yanase2014,maruyama2015}. This is particularly important for $f$-electron system, where on the one hand the spin-orbit interaction caused by heavy atoms can be large, and on the other hand electron correlations in partially filled $f$-electron bands can be very strong. Thus, $f$-electron systems might give rise to a large ME effect. 
The existence of the ME and the inverse ME effect in $f$-electron systems has recently been experimentally demonstrated for the Kondo insulator SmB$_6$\cite{Song2016}.

The previous Fermi liquid analysis how correlations affect the ME effect was however based on the Hubbard model, which is not applicable for $f$-electron systems. In $f$-electron systems, the hybridization between non- or weakly-interacting conduction electrons ($s-$, $p-$, $d-$ orbitals) and strongly interacting $f$-orbitals leads to fascinating phenomena, which are not described by the Hubbard model. While at high temperatures the $f$-electrons are localized and do not participate in the Fermi surface, at low temperatures the Kondo effect leads to the formation of heavy quasi-particles, which are formed by conduction ($c$-)electrons and $f$-electrons. Thus, at low temperatures the $f$-electrons become itinerant and do participate in the Fermi surface\cite{coleman2007}.
This crossover between localized $f$-electrons and itinerant $f$-electrons when the temperature is decreased, the Kondo effect, and the resulting heavy quasi-particles are not included in the previous theoretical works. 

The aim of this paper is to analyze the ME effect, particular the Edelstein effect, in strongly correlated noncentrosymmetric $f$-electron systems such as CeRhSi$_3$, CeIrSi$_3$, or CePt$_3$Si. By using dynamical mean field theory (DMFT), we fully include the Kondo effect and thus the formation of heavy quasi particles and the crossover between localized and itinerant $f$-electrons. Furthermore, the combination of DMFT with the numerical renormalization group (NRG) enables us to calculate transport properties with high accuracy using  real-frequency Green's functions without the need of an analytic continuation.
 
The main results can be summarized as follows: (i) The ME effect can be strongly enhanced in $f$-electron systems and exhibits a maximum at the crossover temperature between localized and itinerant $f$-electrons. This enhancement is beyond Fermi liquid theory. (ii) The enhancement of the ME effect originates from a  coupling between the $c$-electrons and the localized $f$-electrons which generates a momentum dependent spin polarization of the $c$-electrons even at high temperatures, above the formation of heavy quasi-particles. The spin polarization of the $c$-electrons is thereby generated by a virtual hopping between a $c$-electron orbital  and an $f$-electron orbital. Thus, the main contribution to the  enhancement of the ME effect comes from the $c$-electrons.
(iii) Besides the intra-orbital Rashba spin-orbit interaction within the $f$-electron band, the inter-orbital Rashba spin-orbit interaction between $c$-electrons and $f$-electrons is significant for a large ME effect.


\section{Model and method}
\begin{figure}[t]
\begin{center}
  \includegraphics[width=0.99\linewidth]{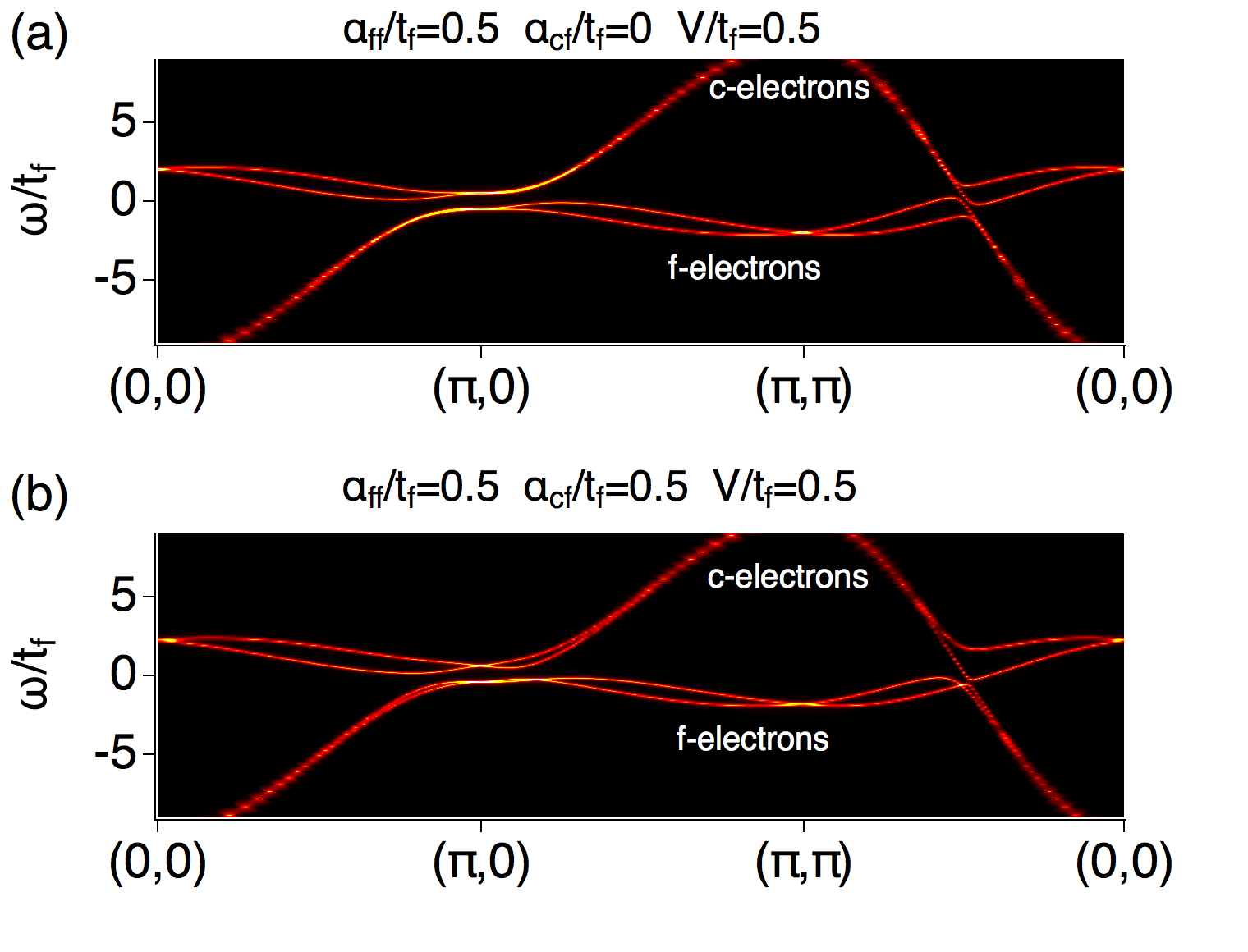}\\
  \end{center}
\caption{Noninteracting momentum resolved spectral functions for (a) $\alpha_{ff}/t_{f}=0.5$, $\alpha_{cf}/t_{f}=0$, $V/t_{f}=0.5$ and (b) $\alpha_{ff}/t_{f}=0.5$, $\alpha_{cf}/t_{f}=0.5$, $V/t_{f}=0.5$.
\label{Fig1}}
\end{figure}
To analyze the ME effect in $f$-electron systems with antisymmetric spin-orbit interaction, we use a periodic Anderson model, which consists of one $c$-electron band and one $f$-electron band and include a local density-density interaction into the $f$ electron band.
Besides a local $c$-$f$ hybridization, we include the intra-orbital  Rashba spin-orbit  interaction within the $f$-electron band, and the inter-orbital Rashba spin-orbit  interaction between $c$-electrons and $f$-electrons. 
Due to the hybridization between strongly correlated $f$-electrons and $c$-electrons, this model includes all essential ingredients necessary to describe heavy fermion behavior. Furthermore, the inclusion of the intra-orbital and inter-orbital Rashba spin-orbit interaction, which have been derived for CePt$_3$Si, reflects the situation of a system without inversion symmetry\cite{Yanase2008} which will lead to the emergence of ME effect.

The Hamiltonian can be split in a single-electron part, $H_{\vec{k}}$, and the interaction part, $H_U$, so that $H=H_{\vec{k}}+H_U$.  $H_{\vec{k}}$ reads
\begin{widetext}
\begin{equation}
H_{\vec{k}}=\sum_{\vec{k}}
\begin{pmatrix}c^\dagger_{k,\uparrow}\\
c^\dagger_{k,\downarrow}\\
f^\dagger_{k,\uparrow}\\
f^\dagger_{k,\downarrow}
\end{pmatrix}\tran
\begin{pmatrix}
t_c(\cos k_x+\cos k_y)&0&V&\alpha_{cf}(\sin k_y+i\sin k_x)\\
0&t_c(\cos k_x+\cos k_y)&\alpha_{cf}(\sin k_y-i\sin k_x)&V\\
V&\alpha_{cf}(\sin k_y+i\sin k_x)&t_f(\cos k_x+\cos k_y)&\alpha_{ff}(\sin k_y+i\sin k_x)\\
\alpha_{cf}(\sin k_y-i\sin k_x)&V&\alpha_{ff}(\sin k_y-i\sin k_x)&t_f(\cos k_x+\cos k_y)
\end{pmatrix}
\begin{pmatrix}c_{k,\uparrow}\\
c_{k,\downarrow}\\
f_{k,\uparrow}\\
f_{k,\downarrow}
\end{pmatrix},
\end{equation}
\end{widetext}
where $c^\dagger_{k,\sigma}$ and $f^\dagger_{k,\sigma}$ create a $c$-electron and an $f$-electron with momentum $k$ and spin projection $\{\uparrow,\downarrow\}$, respectively. Our model includes a spin-independent hopping for the $c$- and $f$-electron with amplitude $t_c$ and $t_f$. We fix the hopping to $t_f=-0.2t_c$. For simplicity we assume a band structure corresponding to a square lattice, but note that our results do not depend on the exact band structure. $V$ is the local hybridization between $c$-electrons and 
$f$-electrons, $\alpha_{cf}$ the inter-orbital Rashba interaction between $c$-electrons and $f$-electrons, $\alpha_{ff}$ the intra-orbital Rashba interaction within the 
$f$-electron band.
The local density-density interaction within the $f$-electron band reads
\begin{equation}
H_U=U\sum_i n^f_{i,\uparrow} n^f_{i,\downarrow}.
\end{equation}
The non-interacting spectrum is shown in Fig. \ref{Fig1} for two parameter sets. Clearly visible are the $c$-electron and $f$-electron bands, which hybridize close to the Fermi energy.  The main difference between these two parameter sets is the band splitting due to the Rashba interaction close to the Fermi energy.  Furthermore, it is important to note that the particle-hole symmetry is generally broken when $V$, $\alpha_{cf}$, and $\alpha_{ff}$ are all nonzero.

To analyze transport properties of this system, we solve the Hamiltonian by using the DMFT\cite{Metzner1989,Georges1996,Pruschke1995}. DMFT maps the lattice model onto a quantum impurity model, which is solved self-consistently. DMFT thereby fully includes local fluctuations, but neglects nonlocal fluctuations. 
The neglect of nonlocal fluctuations is the main drawback of DMFT. It thus must be noted that all obtained results are only valid as long as nonlocal fluctuations are small. However,  DMFT has proven to accurately describe heavy-fermion physics as necessary to analyze $f$-electron materials\cite{PhysRevB.61.12799}. For solving the quantum impurity model, we use the NRG\cite{wilson1975,Bulla2008,Peters2006}, which provides real-frequency spectral functions and self-energies  with high accuracy around the Fermi energy for a wide range of interaction parameters and temperatures.

\section{conductivity and magnetoelectric effect}
The electric current, $J_x$, and the polarization, $M_y$,  are related to the electric field, $E_x$, via the conductivity, $\sigma_{xx}$, and the ME coefficient, $\Upsilon_{yx}$, by
\begin{eqnarray}
J_{x}&=&\sigma_{xx}E_{x}\\
M_{y}&=&\Upsilon_{yx}E_x,
\end{eqnarray}
where $\sigma_{xx}$ and $\Upsilon_{yx}$ can be calculated using the Kubo formula\cite{PhysRevB.47.3553,Pruschke1995}.These two equations can be combined to give the spin polarization depending on the electric current,
\begin{equation}
M_{y}=\frac{\Upsilon_{yx}}{\sigma_{xx}}J_x,
\end{equation}
which we will below use to quantize the strength of the ME effect.

After having obtained self-consistent self-energies using the DMFT, we use the Kubo formula to calculate the conductivity, $\sigma_{xx}(\omega)$, and the ME effect, $\Upsilon_{yx}(\omega)$, which are defined as
\begin{eqnarray}
\sigma_{xx}(\omega)&=&\frac{i}{\omega}\sum_{k,k^\prime}\text{Tr}\langle\langle v_{x} n_k,v_{x}n_{k^\prime}\rangle\rangle(\omega)\\
\Upsilon_{yx}(\omega)&=&\frac{i}{\omega}\sum_{k,k^\prime}\text{Tr}\langle\langle \sigma_y n_k,v_{x}n_{k^\prime}\rangle\rangle(\omega).
\end{eqnarray}
We have set $\hbar=e=\mu_0=1$. The main problem consists of calculating the two-particle Green's functions $\langle\langle v_{x} n_k,v_{x}n_{k^\prime}\rangle\rangle(\omega)$ and $\langle\langle \sigma_y n_k,v_{x}n_{k^\prime}\rangle\rangle(\omega)$, where the difference between the conductivity and the ME effect is the change from the velocity operator, $v_x$, to the Pauli-spin matrix, $\sigma_y$.  We take the same Pauli matrix in the $c$- and $f-$ electron bands setting $g=2$ and remind the reader that all operators represent $4\times 4$ matrices, so that
\begin{equation}
\sigma_y=\begin{pmatrix}0&-i&0&0\\i&0&0&0\\0&0&0&-i\\0&0&i&0\end{pmatrix}
\end{equation}
Because vertex corrections are neglected within the DMFT approximation, the two-particle Green's function reduces to the product of two single-particle Green's functions written in Matsubara frequencies as
\begin{eqnarray}
\sigma_{xx}(i\omega)&=&\frac{1}{\omega}\Pi_{xx}(i\omega)\\
\Upsilon_{yx}(i\omega)&=&\frac{1}{\omega} K_{yx}(i\omega)\\
\Pi_{xx}(i\omega)&=&T\sum_{k}\sum_{i\nu}\text{Tr}\Big[ v_xG_k(i\nu)v_xG_k(i\nu+i\omega)\Big]\\
K_{yx}(i\omega)&=&T\sum_{k}\sum_{i\nu}\text{Tr}\Big[ \sigma_yG_k(i\nu)v_xG_k(i\nu+i\omega)\Big]\\
\end{eqnarray}
where $T$ is the temperature of the system. 

Having a self-consistent solution for the self-energy, these single-particle Green's functions are known and  $\Pi_{xx}(i\omega)$ and $K_{yx}(i\omega)$ could be calculated using Matsubara frequencies, which must be followed by an analytic continuation at the end of the calculation.

However, a significant advantage of combining DMFT with NRG is the availability of real-frequency spectral functions and self-energies. Thus, we can perform the full calculation using real frequencies, which results in a considerable gain of accuracy.
For each component of the Green's function, we can write
\begin{equation}
G_k(z)=\int d\omega \frac{1}{z-\omega}A_k(\omega)
\end{equation}
where $A_k(\omega)=\frac{1}{2\pi i}(G_k^{ret}(\omega)-G_k^{adv}(\omega))$ is the density of states, which is calculated from  the retarded and advanced Green's functions, $G_k^{ret}(\omega)$ and $G_k^{adv}(\omega)$. 

Writing the density of states for all components of the Green's function again as a matrix, we can calculate $\Pi_{xx}(\omega)$ and $K_{yx}(\omega)$ directly on the real-frequency axis. The conductivity $\sigma_{xx}$ and ME effect $\Upsilon_{yx}$ thus become
\begin{widetext}
\begin{eqnarray}
\sigma_{xx}(\omega)&=&\sum_k\int d\omega^\prime \text{Tr} \Big [v_x A(\omega^\prime) v_x A(\omega+\omega^\prime)\Big ]\frac{f_T(\omega^\prime)-f_T(\omega+\omega^\prime)}{\omega}\\
\Upsilon_{yx}(\omega)&=&\sum_k\int d\omega^\prime \text{Tr} \Big [\sigma_y A(\omega^\prime) v_x A(\omega+\omega^\prime)\Big ]\frac{f_T(\omega^\prime)-f_T(\omega+\omega^\prime)}{\omega},
\end{eqnarray}
\end{widetext}
where $f_T(\omega)$ is the Fermi-function for temperature $T$.
Taking the static limit $\omega\rightarrow 0$, we obtain the final result
\begin{eqnarray}
\sigma_{xx}(\omega=0)&=&\sum_k\int d\omega^\prime \text{Tr} \Big [v_x A(\omega^\prime) v_x A(\omega^\prime)\Big ]\frac{d f_T(\omega^\prime)}{d\omega^\prime}\\
\Upsilon_{yx}(\omega=0)&=&\sum_k\int d\omega^\prime \text{Tr} \Big [\sigma_y A(\omega^\prime) v_x A(\omega^\prime)\Big ]\frac{d f_T(\omega^\prime)}{d\omega^\prime}.
\end{eqnarray}
We note that in these results the temperature dependence enters via the Fermi function and $A(\omega)$ which depends on the self-energy calculated self-consistently for a given temperature.


\section{Noninteracting system}
\begin{figure}[t]
\begin{center}
    \includegraphics[width=0.9\linewidth]{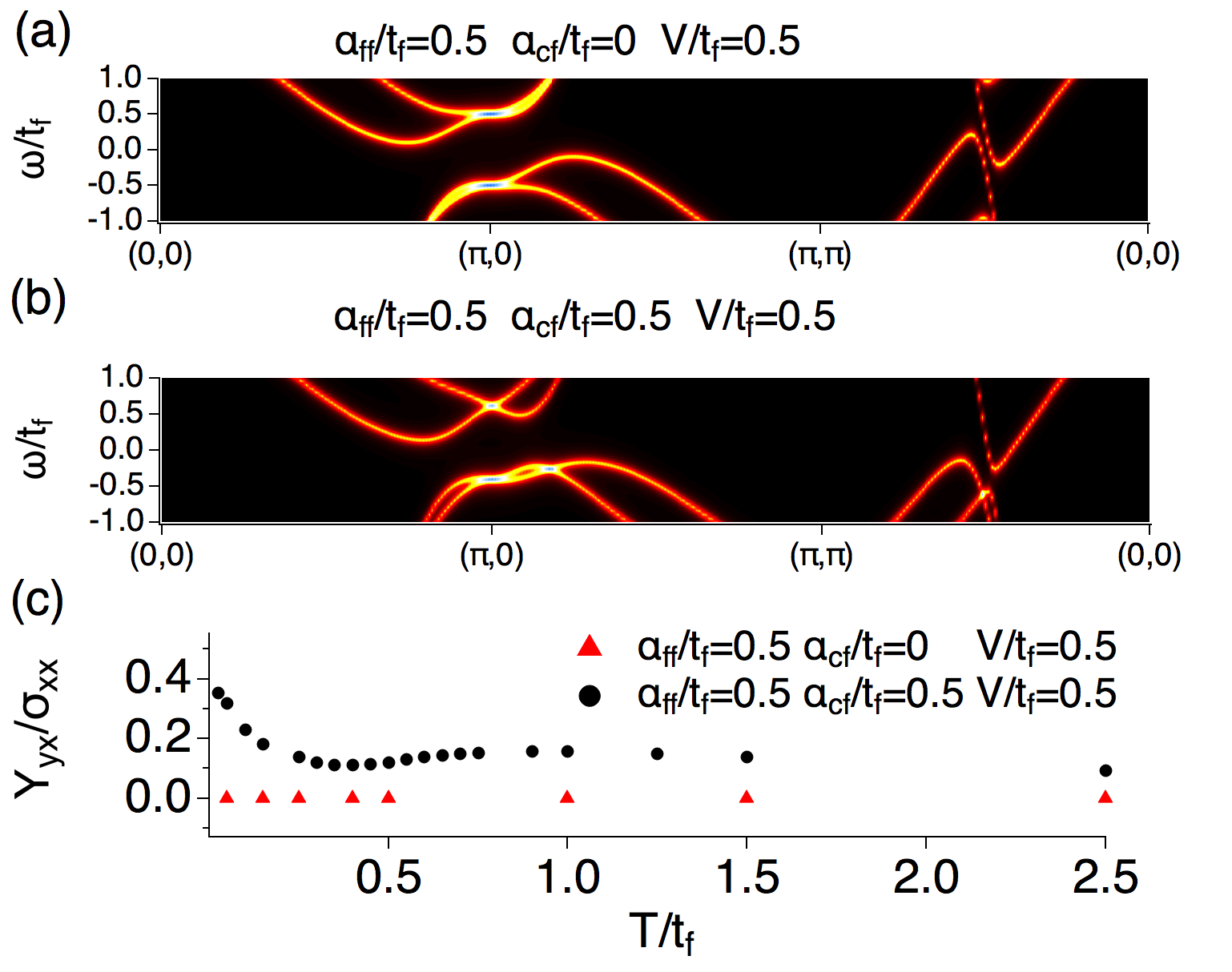}
\end{center}
\caption{(a) and (b) Noninteracting momentum resolved spectral functions. Model parameters are written above each panel. The Fermi energy corresponds to $\omega/t_f=0$. (c) ME effect for the parameter shown in (a)-(b).
\label{Fig2}}
\end{figure}
\begin{figure}[t]
\begin{center}
    \includegraphics[width=0.99\linewidth]{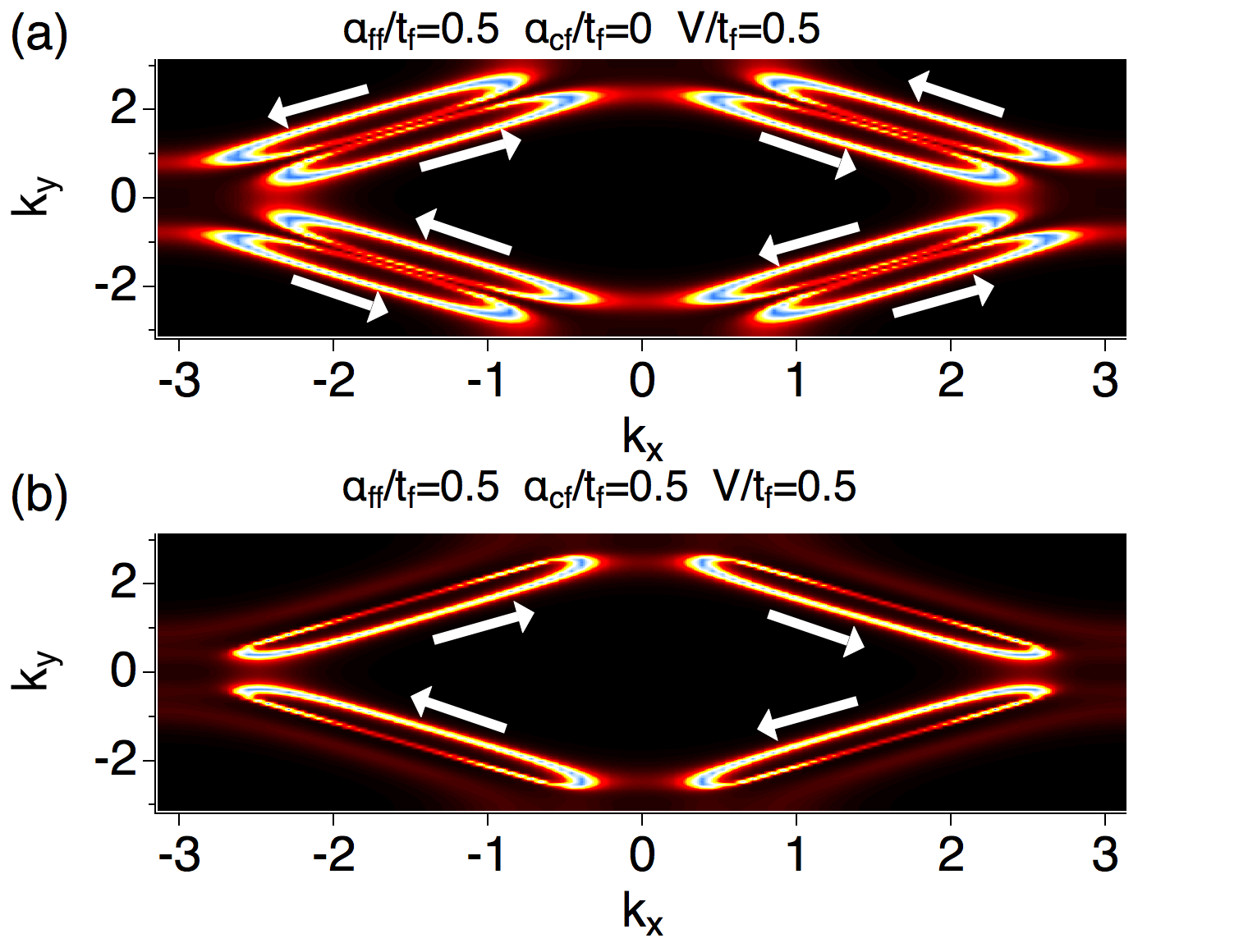}
  \end{center}
\caption{Fermi surfaces of the non-interacting systems shown in Figs. \ref{Fig1} and \ref{Fig2}.
The arrows in the plot indicate the direction of the spin polarization in these bands, which is induced due to the Rashba spin-orbit interaction.
\label{Fig3}}
\end{figure}

To gain some understanding about this model, we firstly  show results for the noninteracting system in Fig. \ref{Fig2}. This will help to clarify the effect of the Coulomb interaction below.
Close to the Fermi energy, shown in Fig. \ref{Fig2}, the visible bands are composed of hybridized  $f$-electrons and $c$-electrons. Both systems are metallic with a spin-split Fermi surface.
The chemical potential is adjusted in 
both systems, so that the system is half-filled, $n_f=n_c=1$. We note that while Fig. \ref{Fig2}(a) corresponds to a particle-hole symmetric system, the particle-hole symmetry is broken in Fig. \ref{Fig2}(b).
Figure \ref{Fig2}(c) shows the results for $\Upsilon_{yx}/\sigma_{xx}$ for the parameter sets in (a) and (b).

For the system with $\alpha_{cf}=0$, shown in Fig. \ref{Fig2}(a), the ME effect disappears at half-filling. Due to the Rashba interaction, two bands with opposite spin polarization cut the Fermi energy. Therefore,  the particle-hole symmetry, which is conserved in this system, results in a perfect cancellation of the contributions of these bands to the ME effect. To verify this statement, we show the Fermi surface of this parameter set in Fig. \ref{Fig3}(a) and include the spin polarization of each band. Clearly visible is the appearance of bands with identical shape but opposite spin polarization.  

On the other hand, the system including all three parameters ($V\neq0$ , $\alpha_{ff}\neq0$ , $\alpha_{cf}\neq0$), shown in Fig. \ref{Fig1}(b) and Fig. \ref{Fig2}(b), has a finite ME effect. The inclusion of $\alpha_{cf}$ breaks the particle-hole symmetry and favours bands with equal spin-polarization close to the Fermi energy. Therefore, the contributions to the ME effect from different bands at the Fermi energy do not completely cancel. The Fermi surface including the spin polarization is shown in Fig.\ref{Fig3}(b). Due to the breaking of particle-hole symmetry, bands with opposite spin polarization have vanished from the Fermi surface.

These results demonstrate the importance of the inter-orbital Rashba interaction for the ME effect which naturally arises in noncentrosymmetric $f$-electron systems and leads to bands with spin polarization into the same direction. We note that the exact cancellation for the system with $\alpha_{cf}=0$ only holds for the half-filled situation. The ME effect becomes finite, when doping the system away from half-filling.


\section{Interacting system}
 \begin{figure}[t]
\begin{center}
  \includegraphics[width=0.9\linewidth]{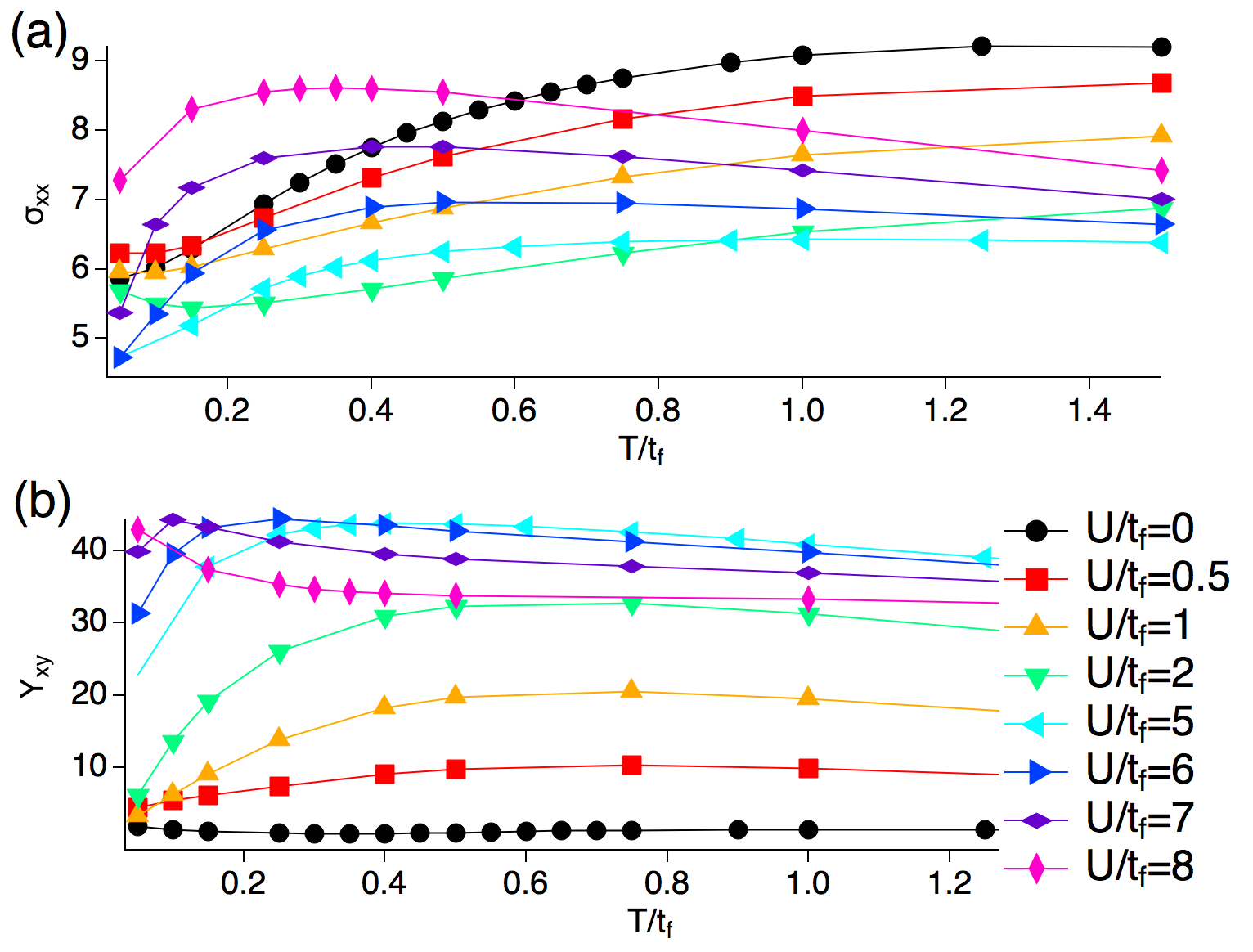}
  \end{center}
\caption{Conductivity $\sigma_{xx}$ (panel a) and magnetoelectric effect $\Upsilon_{xy}$ (panel b) for $\alpha_{ff}/t_{f}=\alpha_{cf}/t_{f}=V/t_{f}=0.5$ and different interaction strengths and temperatures. 
\label{Fig4}}
\end{figure}

 \begin{figure}[t]
\begin{center}
  \includegraphics[width=0.9\linewidth]{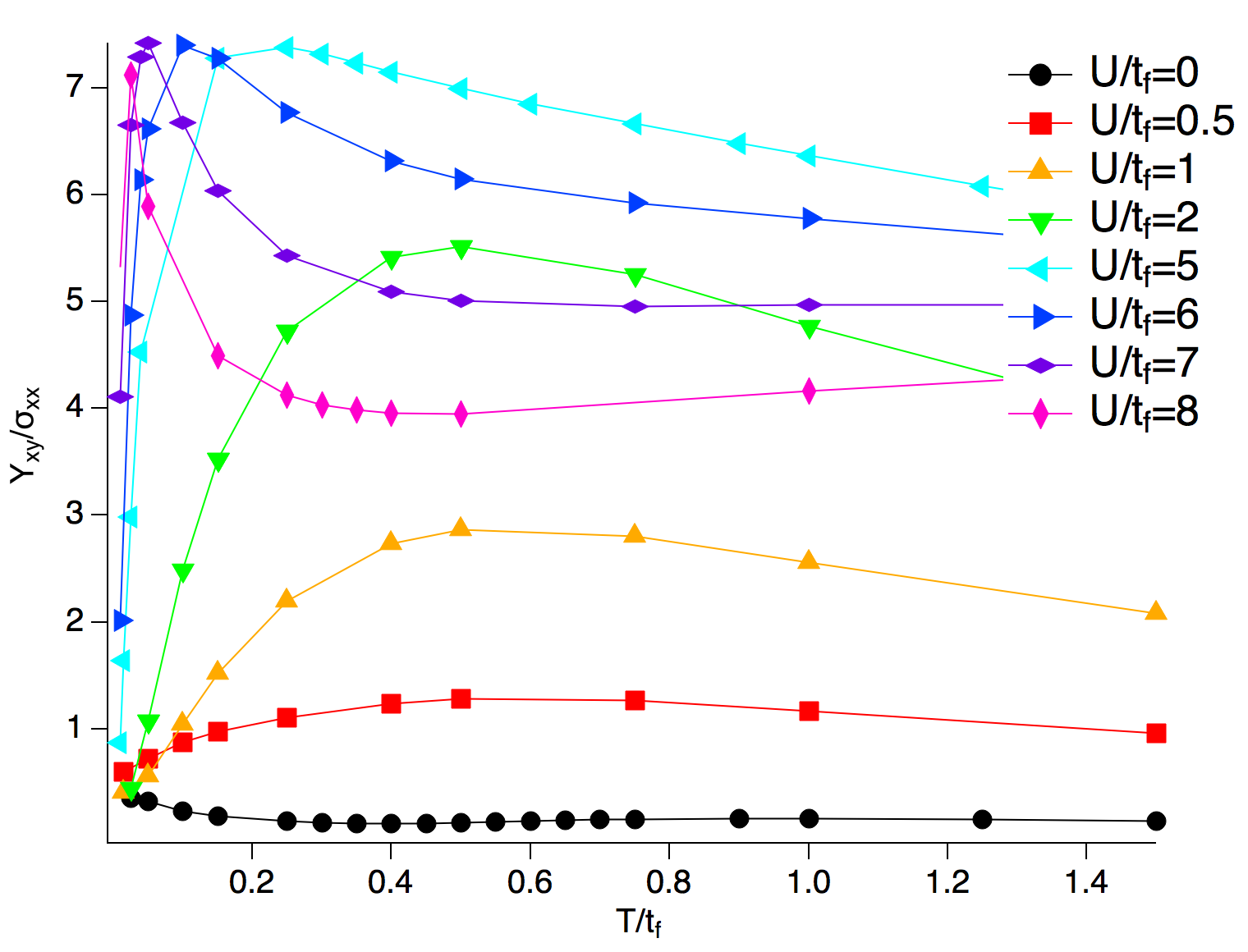}
  \end{center}
\caption{$\Upsilon_{xy}/\sigma_{xx}$ for $\alpha_{ff}/t_{f}=\alpha_{cf}/t_{f}=V/t_{f}=0.5$ for different interaction strengths and temperatures. 
\label{Fig5}}
\end{figure}
We next turn our attention to the interacting system. 
Because the system with $\alpha_{cf}=0$ which preserves particle-hole symmetry 
 can be regarded as a special situation, 
we will focus from now on the metallic system with $V=\alpha_{ff}=\alpha_{cf}=0.5t_f$. The chemical potential is adjusted in all calculations so that the system remains half-filled. 
We note that the qualitative behavior shown here does not depend on the filling of the conduction electrons. 

Let us start the analysis by showing separately the conductivity, $\sigma_{xx}$, and ME effect, $\Upsilon_{xy}$, for different interaction strengths and temperatures, see Fig. \ref{Fig4}. For the noninteracting system, the conductivity decreases with decreasing temperature in the shown temperature range due to the hybridization between $c$- and $f$-electrons which gaps out parts of the Fermi surface. With increasing interaction strength, the conductivity develops a peak at finite temperature. Overall, the conductivity exhibits only a moderate interaction dependence for the shown temperatures. On the other hand, the ME effect shown in Fig. \ref{Fig4}(b) is small at weak-coupling, but strongly increases with increasing interaction strength. It develops a peak at a finite temperature. The peak position is at a slightly  smaller temperature than that in the conductivity. At very low temperature the ME effect strongly decreases again.

Figure \ref{Fig5} shows the temperature dependent ratio of ME effect and conductivity
for  different interaction strengths, which can be measured in experiment. 
We see that even  a weak  interaction in the $f$-orbital, $U/t_f=0.5$, enhances the ME effect, particularly at high temperatures. Comparing with Fig. \ref{Fig4}, it becomes clear that this enhancement is due to the enhancement of the ME effect, and not due to a strong change in the conductivity.
Increasing the interaction further, we find a significant enhancement of the ME effect and a clear peak in the temperature dependence. While the height of this peak increases with the interaction strength for $U/t_f<5$, it becomes constant when further increasing the interaction.  The temperature of this peak  decreases monotonically with increasing interaction and can be identified as the crossover temperature between localized and itinerant $f$-electrons.

Comparing to the ME effect of the noninteracting system, 
we observe that the maximum value is more than ten times enhanced by the correlations. However, the enhancement is even more dramatic at high temperatures, where we find a ME effect nearly $40$ times the noninteracting value.
Thus, our results suggest to look at the ME effect in heavy fermion systems above their coherence temperature.
This enhancement would be most useful for spintronics application at room temperature.

\begin{figure}[t]
\begin{center}
  \includegraphics[width=0.9\linewidth]{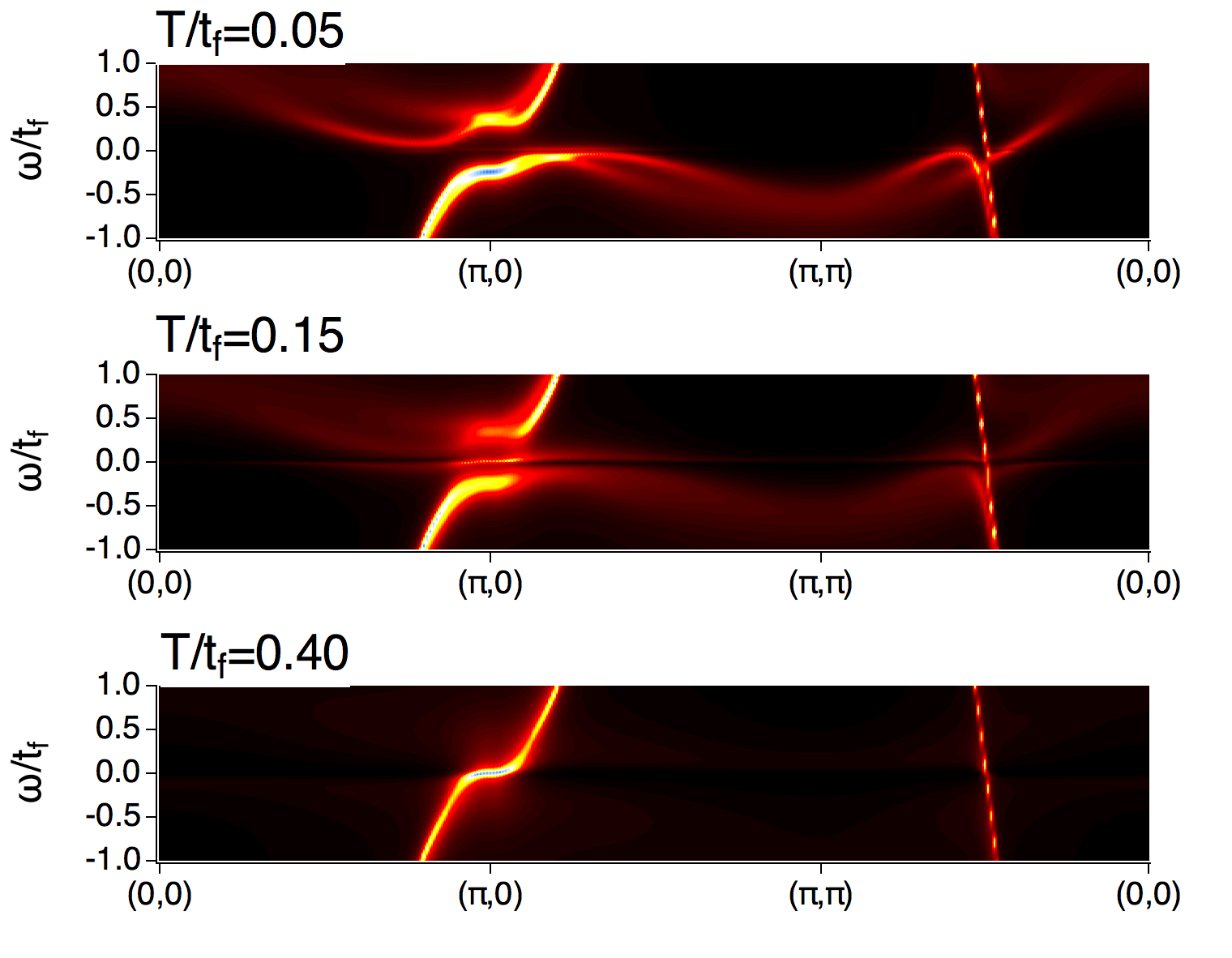}
\end{center}
\caption{Momentum resolved spectral functions for $U/t_f=5$, $\alpha_{ff}/t_{f}=\alpha_{cf}/t_{f}=V/t_{f}=0.5$ and different temperatures.
\label{Fig6}}
\end{figure}
\begin{figure}[t]
\begin{center}
  \includegraphics[width=0.9\linewidth]{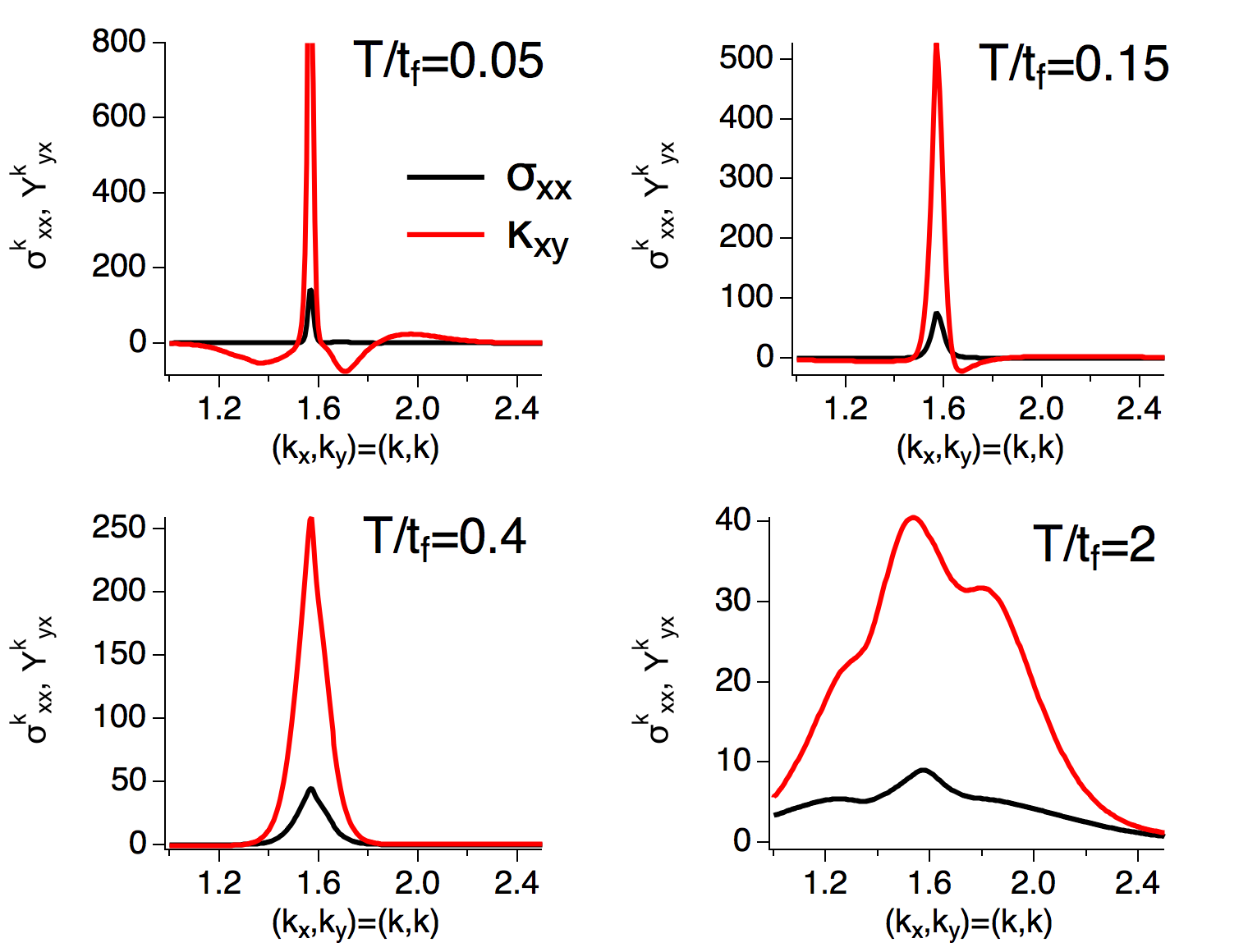}
\end{center}
\caption{Momentum resolved $\sigma^k_{xx}$ and $\Upsilon^k_{yx}$ along the diagonal in the Brillouin zone for $U/t_f=5$, $\alpha_{ff}/t_{f}=\alpha_{cf}/t_{f}=V/t_{f}=0.5$ and different temperatures.
\label{Fig7}}
\end{figure}
\begin{figure*}[t]
\begin{center}
  \includegraphics[width=0.3\linewidth]{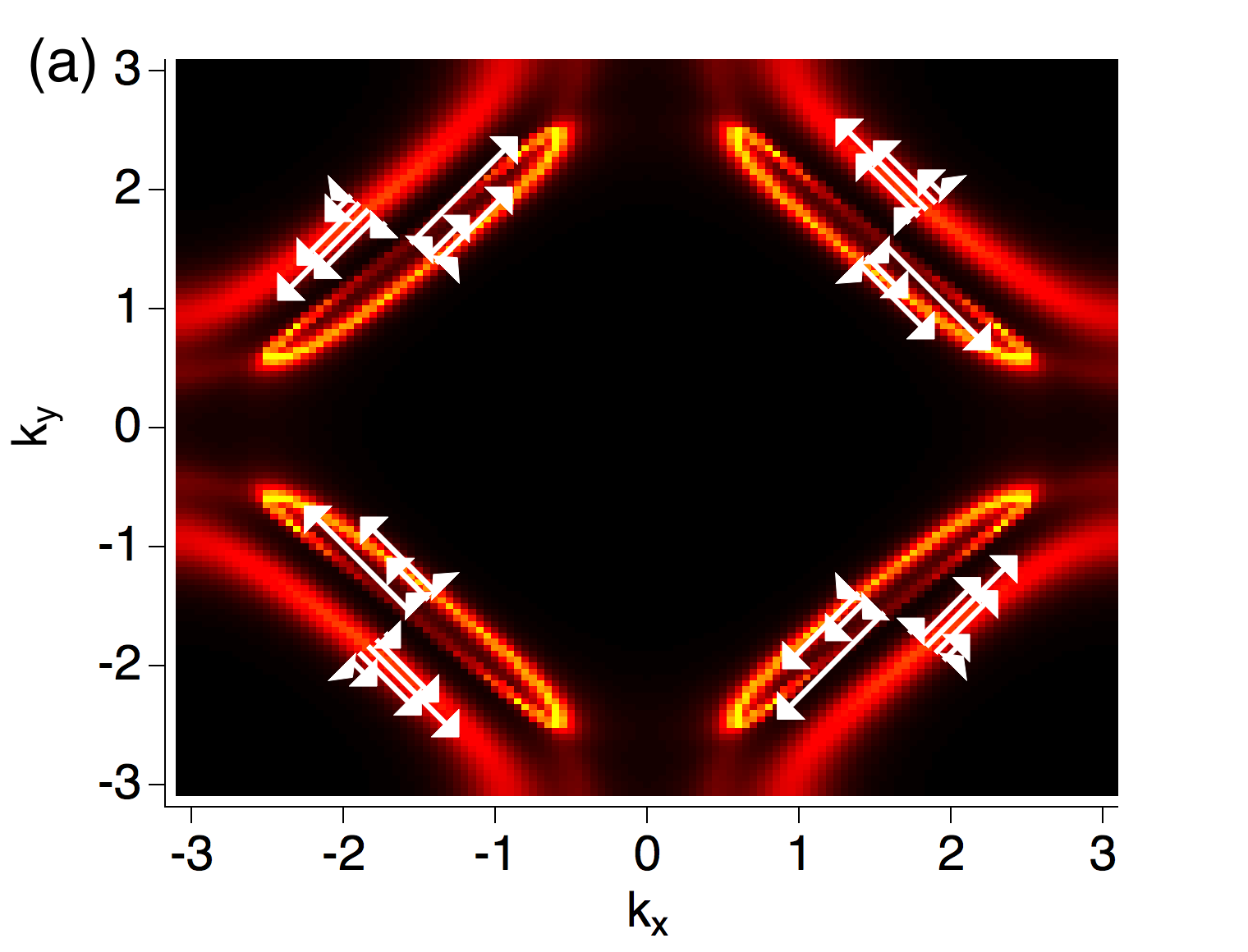}
    \includegraphics[width=0.3\linewidth]{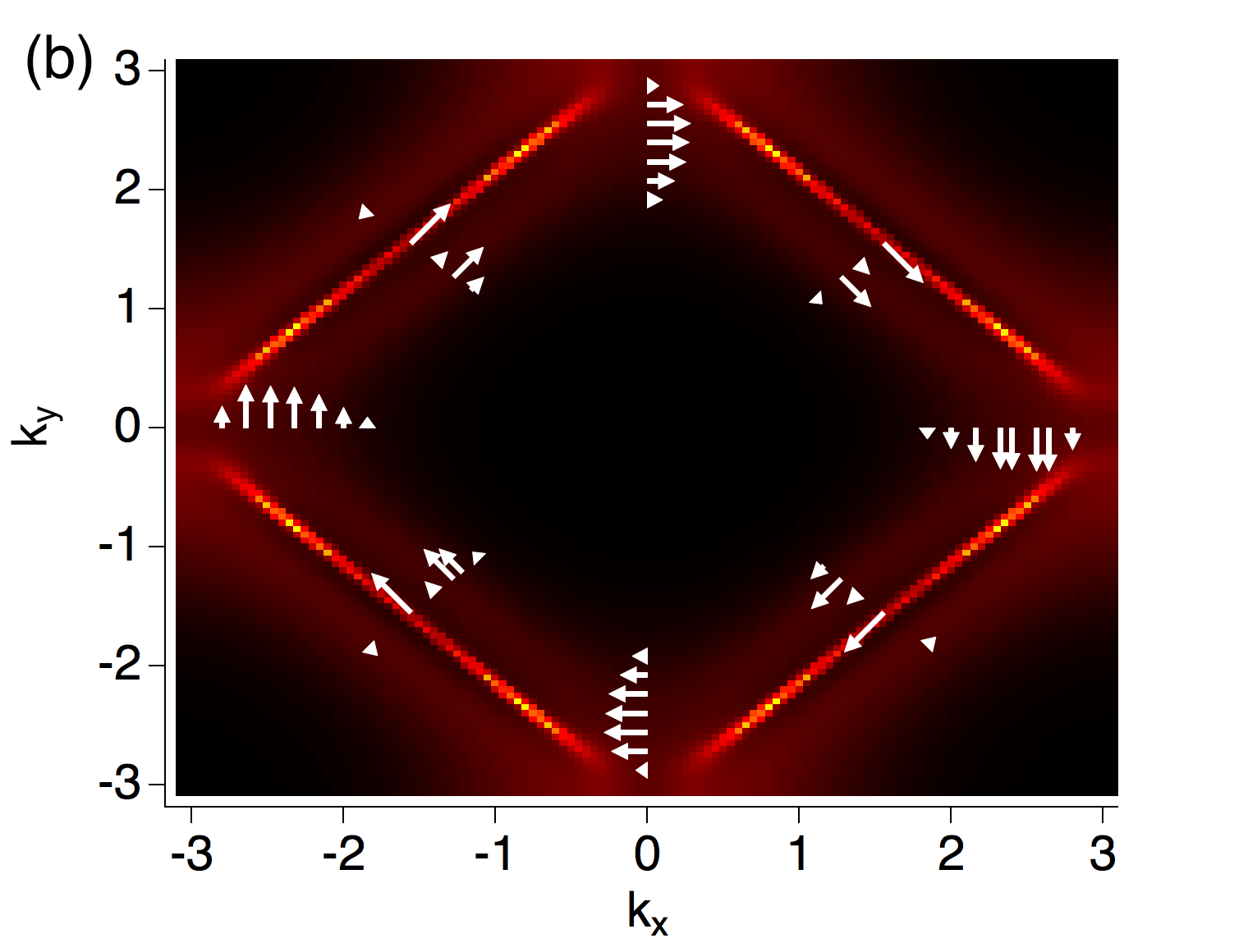}
    \includegraphics[width=0.3\linewidth]{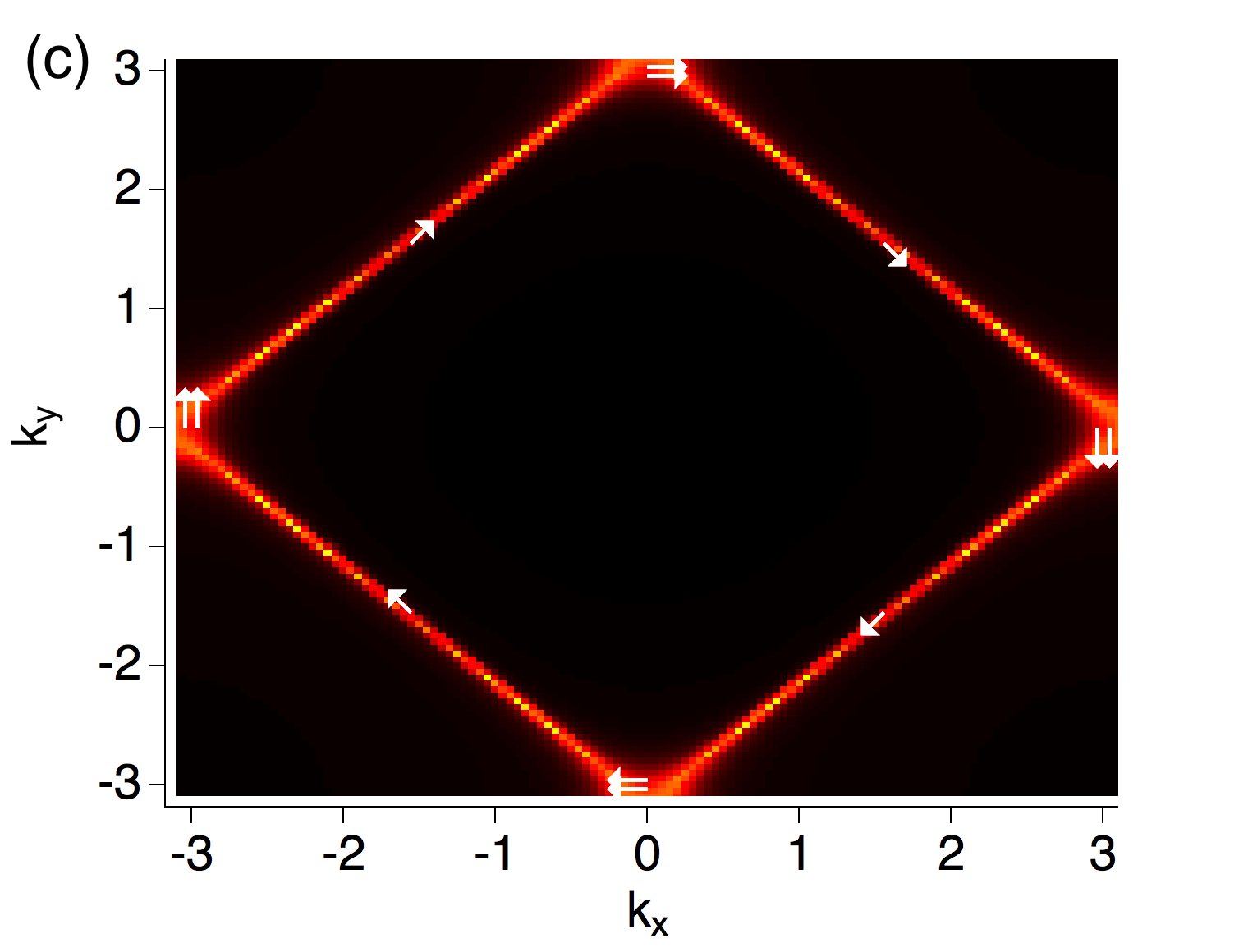}
  \end{center}
\caption{Cuts of the momentum dependent density of states through the Brillouin zone at the Fermi energy. The color plot denotes the density of states at the fixed energy with a maximum intensity for yellow. The white arrows denote the calculated spin polarization. The parameters are $\alpha_{ff}/t_f=0.5$ $\alpha_{cf}/t_f=0.5$ $V/t_f=0.5$
and $U/t_f=5$. (a):$T/t_f=0.01$ (b): $T/t_f=0.15$ (c): $T/t_f=0.4$.
\label{Fig8}}
\end{figure*}

To understand the mechanism behind this enhancement, we show momentum resolved spectral function for $U/t_f=5$ in Fig. \ref{Fig6}. The depicted temperatures correspond to a low temperature, where the ME effect is small ($T/t_f=0.05$), a temperature shortly below the peak of the ME effect ($T/t_f=0.15$), and a temperature shortly above the peak ($T/t_f=0.4$). The spectral function at temperatures above the peak includes only 
$c$-electrons;  due to a strong peak in the imaginary part of the $f$-electron self-energy, the $f$-electrons are completely localized and thus absent from the spectral function. Heavy quasi-particles are not formed at these temperatures. Because there is no Rashba spin-orbit interaction acting within the conduction band, a spin splitting of the conduction band is not observed. 
At temperatures, shortly below the peak of the ME effect, we observe the appearance of the $f$-electron band within the spectrum. 
The $f$-electrons become itinerant at this temperature and begin to form heavy-quasi particles, which are observable as flat band at the Fermi energy.
 This proves that the peak of the ME effect is related to the coherence temperature of the system. Finally, at very low temperature, we find coherent heavy quasi-particles around the Fermi energy. The spectrum looks similar to the noninteracting spectrum with renormalized energies.

To elucidate the reason for the enhancement at high temperatures, we show the 
summand of the momentum integration for the conductivity ($\sigma^k_{xx}=\int d\omega^\prime \text{Tr} \Big [v_x A_k(\omega^\prime) v_x A_k(\omega^\prime)\Big ]\frac{d f_T(\omega^\prime)}{d\omega^\prime}$) and the ME effect ($\Upsilon^k_{yx}=\int d\omega^\prime \text{Tr} \Big [\sigma_y A_k(\omega^\prime) v_x A_k(\omega^\prime)\Big ]\frac{d f_T(\omega^\prime)}{d\omega^\prime}$) along the diagonal of the Brillouin zone in Fig. \ref{Fig7}. 
 The conductivity and the ME effect as shown in the previous figures correspond to the momentum integral of these functions over the whole Brillouin zone. The summand for the conductivity is always positive. Its amplitude around the Fermi momentum $(\pi/2,\pi/2)$ is increasing with decreasing temperature due to an increased lifetime, while the width of the peak decreases at the same time. The summand of the ME effect, on the other hand, shows a more interesting behavior. At low temperature, $T/t_f=0.05$, it shows positive as well as negative contributions. 
 
 The existence of positive and negative contributions can also be immediately  understood from the Fermi surface of the system ($\alpha_{ff}/t_f=0.5$, $\alpha_{cf}/t_f=0.5$, $V/t_f=0.5$, $U/t_f=5$) including the spin polarization  shown in Fig. \ref{Fig8} for  three different temperatures. Figure \ref{Fig8}(a) shows a low temperature, where the $f$-electrons are itinerant. We observe a complicated Fermi surface made up of several bands. Furthermore, we observe that these bands have opposite spin polarization, which results in a cancellation of the ME effect at this temperature. This cancellation is indeed responsible for the suppression of the ME effect in most metallic systems.
 
  However, the situation is very different at high temperatures. In Fig. \ref{Fig7}, we observe that the negative contribution to the ME effect vanishes with increasing temperature. At temperatures above the coherence temperature, we only find positive contributions. 
  In Fig. \ref{Fig8}(b), the $f$-electron bands become incoherent and are blurred in the density of states at the coherence temperature. Nevertheless, the $f$-electrons still contribute to the spin polarization. Thus, the spin polarization includes momentum regions with opposite direction. Finally, Fig. \ref{Fig8}(c) shows a temperature where the $f$-electrons are localized and thus are absent from the spectrum. The calculated spin polarization only includes the clockwise direction. There are only positive contributions to the ME effect at this temperature, see $T/t_f=0.4$ in Fig. \ref{Fig7}.
 We thus conclude that the ME effect becomes large because of an absence of cancellation above the crossover temperature between itinerant and localized $f$-electrons.

Because the $f$-electrons are localized, the ME effect above the coherence temperature is solely generated by the $c$-electrons.
It is rather remarkable that the $c$-electrons contribute to the ME effect when the $f$-electrons are absent from the spectrum, although there is no direct Rashba interaction within the $c$-orbitals.
This fact can be understood in the following way: In a virtual  
 process, a $c$-electron can hop onto an $f$-orbital and return to a $c$-electron orbital. 
 This hopping process involves the inter-orbital Rashba spin-orbit interaction and thus will lead  
 to a term describing a spin-dependent coupling, 
  which can generate a spin polarization within the $c$-electron band. 
  Thus, the dependence of the ME effect on the interaction strength and the temperature can be understood as an interplay between the localization of $f$-electrons and a virtual hopping of $c$-electrons on $f$-electron orbitals.  At high temperature, $f$-electrons are localized due to the Coulomb interaction. Thus, the ME effect arises due to polarized $c$-electrons and it is large because of an absence of cancellation. With lowering the temperature, the ME effect firstly increases until the coherence temperature of the material is reached. At this temperature the $f$-electrons become itinerant. 
 At lower temperatures, the material is described by a  renormalized band structure of the noninteracting one. Thus, the ME effect is small due to cancellation effects.
  The coherence temperature, where the $f$-electrons change from localized to itinerant, decreases thereby strongly with increasing Coulomb interaction.

\begin{figure}[t]
\begin{center}
  \includegraphics[width=1\linewidth]{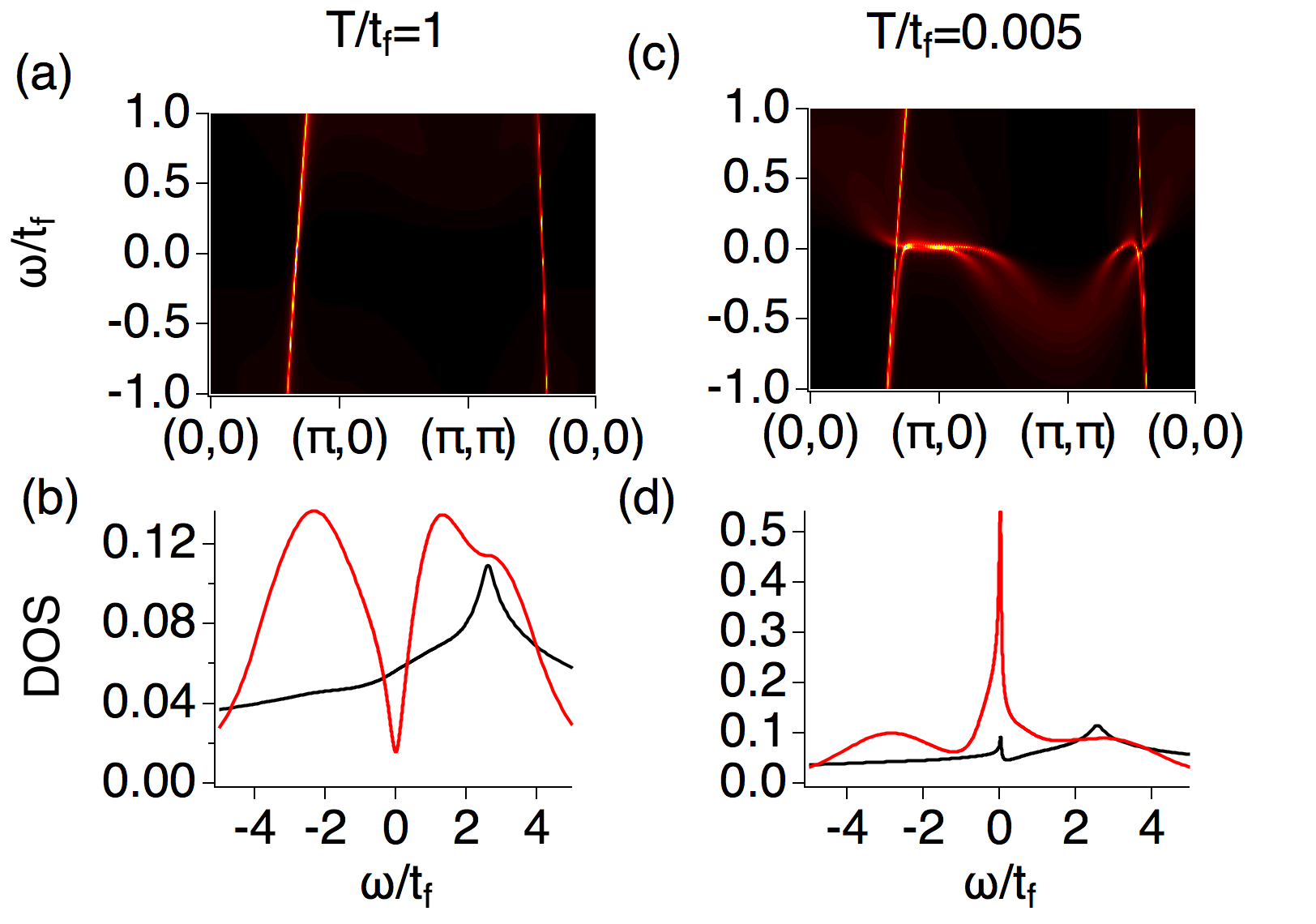}
\end{center}
\caption{System with $\alpha_{ff}/t_f=0.5$, $\alpha_{cf}/t_f=0.5$, $V/t_f=0.5$ and $U/t_f=5$ at $T/t_f=1$ (left panels) and $T/t_f=0.005$ (right panels). The filling of the $c$-electron band is $n_c=0.6$. The $f$-electron band is half-filled. (a) Momentum-resolved spectral function for $T/t_f=1$. (b) Local density of states (DOS) for $T/t_f=1$. The black (red) lines correspond to the $c$- ($f$-) electrons. (c)  Momentum-resolved spectral function for $T/t_f=0.005$. (d) Local density of states (DOS) for $T/t_f=0.005$.
\label{Fig9}}
\end{figure}
\begin{figure}[t]
\begin{center}
  \includegraphics[width=1\linewidth]{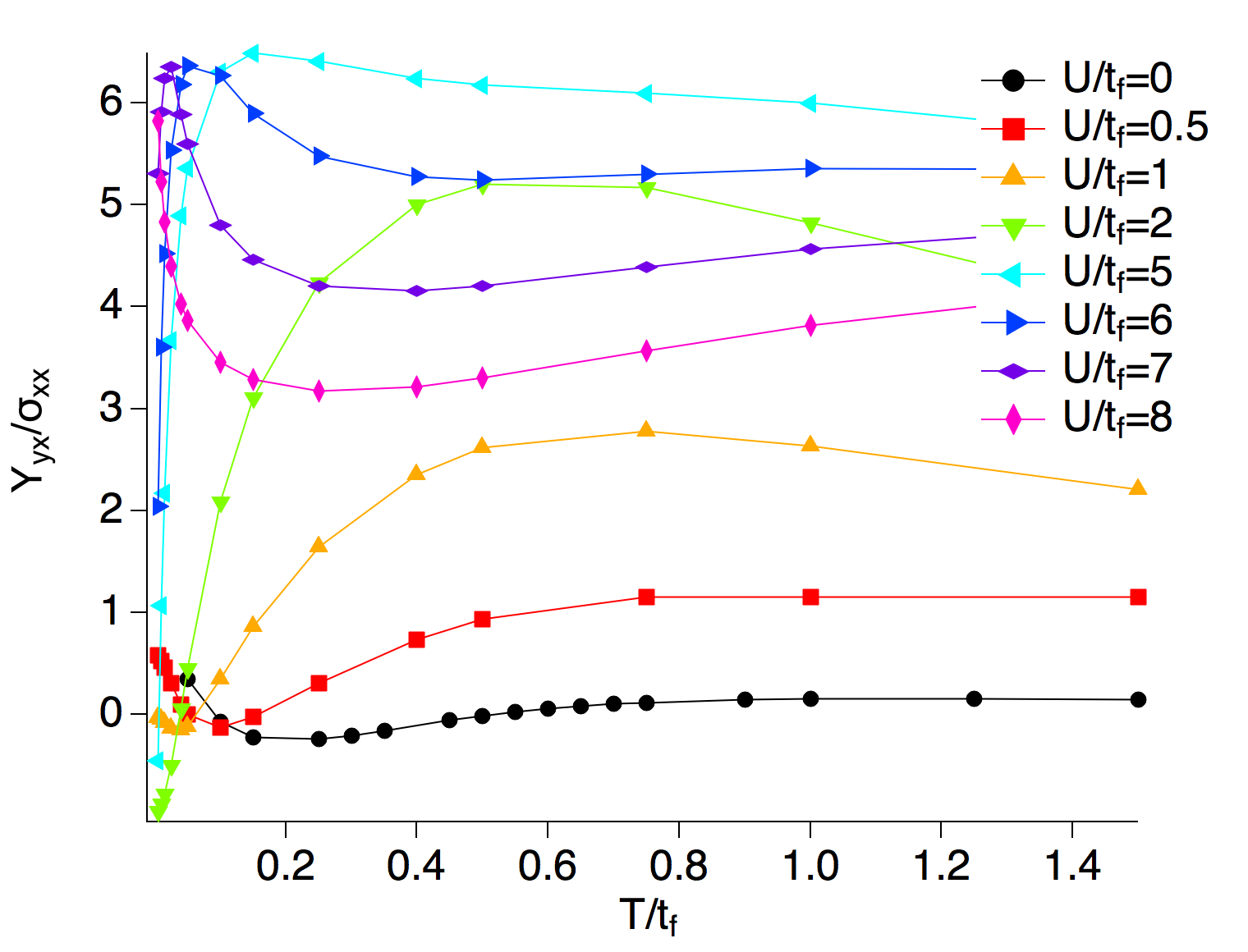}
\end{center}
\caption{ME effect $\Upsilon_{yx}/\sigma_{xx}$ for $\alpha_{ff}/t_f=0.5$, $\alpha_{cf}/t_f=0.5$, $V/t_f=0.5$ and filling of the conduction electrons $n_c=0.6$.
\label{Fig10}}
\end{figure}

Finally, before concluding this paper, we want to shortly address the situation for a hole-doped system.
Up to now, we have focused on a half-filled system, which might be regarded as a special situation, although the particle-hole symmetry is broken for $\alpha_{ff}/t_f=0.5$, $\alpha_{cf}/t_f=0.5$, $V/t_f=0.5$.
We here demonstrate that the enhancement of the ME effect does not depend on the filling of the $c$-electron band.

Figure \ref{Fig9} shows the momentum resolved and local DOS of the system with $\alpha_{ff}/t_f=0.5$, $\alpha_{cf}/t_f=0.5$, $V/t_f=0.5$ and interaction strength $U/t_f=5$ at $T/t_f=1$ and $T/t_f=0.005$. While the $f$-electron band is half-filled, the $c$-electron band has a filling $n_c=0.6$.
We observe for the doped system qualitatively the same physics as at half filling. For high temperature, $T/t_f=1$, the $f$-electrons are localized and thus absent from the Fermi energy in the momentum resolved spectral function and local DOS. The conduction electrons show a spectrum corresponding to noninteracting electrons on a square lattice. 

At low temperatues, $T/t_f=0.005$, the $f$-electrons become coherent and form heavy quasi-particles together with the $c$-electrons. The $f$-electrons form a peak in the density of states at the Fermi energy. This is exactly the same physics as described above for the half-filled system. 

It is thus not surprising to find qualitatively similar behavior for the ME effect shown in Fig. \ref{Fig10}. The ME effect is enhanced for the interacting system
and shows a clear peak, which can be identified again as the transition between localized and itinerant $f$-electrons.

\section{Conclusions}
We have demonstrated that the ME effect can be strongly enhanced in $f$-electron systems showing a peak at the coherence temperature, where the $f$-electrons change from itinerant to localized behavior. 
Above the coherence temperature, where a strong peak in the imaginary part of the $f$-electron self-energy is formed, the Fermi liquid theory breaks down, and a momentum-dependent spin polarization of the $c$-electrons is created, which causes the large ME effect.
Remarkably, a cancellation of the ME effect due to spin-split bands with different polarization is absent at this temperature, which is the main reason for the enhancement. 
Thus, our results suggest to look at the ME effect in noncentrosymmetric $f$-electron systems such as CeRhSi$_3$, CeIrSi$_3$, or CePt$_3$Si above their coherence temperature.  The coherence temperature, as defined in our calculation, can be determined from experiment by the peak position of the magnetic contribution to the resistivity. For CeRhSi$_3$ and CeIrSi$_3$ this peak can be observed at approximately $T_c=100$K\cite{Muro1998} and for  for CePt$_3$Si at $T_c=80$K\cite{PhysRevLett.92.027003}. The spin-orbit interaction in CePt$_3$Si has been estimated from first principle calculation to $50$meV-$200$meV \cite{PhysRevB.69.094514}. If we assume the strength of the spin-orbit coupling to be $100$meV in our calculations, $t_f$ will also be $100$meV. Our calculations with $U/t_f=6$ and $U/t_f=7$ would then correspond to coherence temperatures of $T_c=120$K and $T_c=60$K, respectively. The enhancement of the ME effect at room temperature due to interaction effects would be approximately $40$ for these calculations. Thus, our results suggest that these noncentrosymmetric $f$-electron materials might have a large ME effect even at room temperature, which would be most significant for spintronics applications.

\begin{acknowledgments}
We thank R. Takashima for helpful discussion.
This work was supported by a Grant-in-Aid for Scientific Research on
Innovative Areas "J-Physics" (JP15H05884) and "Topological
Materials Science" (JP16H00991) from the Japan Society for the Promotion of Science (JSPS), and by JSPS KAKENHI Grants (Numbers JP15K05164 and
JP15H05745). Computer simulations were performed on the "Hokusai" supercomputer in RIKEN and the supercomputer of  the Institute for Solid State Physics (ISSP) in Japan.  \end{acknowledgments}
\bibliography{paper}

\end{document}